%% file: main.tex
\documentclass[sigconf]{acmart}
\makeatletter
\def\@ACM@copyright@check@cc{}
\makeatother
\setcopyright{cc}
\input{ExSitu/paper-preamble}

\begin{document}

\input{ExSitu/paper-info}

\maketitle

\input{sections/1_Introduction}

\input{sections/2_InteractionApproaches}
\input{sections/3_RolesOfAI}
\input{sections/4_WSproposal}

\bibliographystyle{formalities/ACM-Reference-Format}
\bibliography{main.bib}

\end{document}

%% file: ExSitu/paper-preamble.tex
\usepackage{array}
\usepackage{tabularx}
\usepackage{ltxtable}
\usepackage{graphicx}
\usepackage{booktabs}
\usepackage{makecell} 
\usepackage{stfloats} 
\usepackage{xspace} 
\usepackage{subcaption}

\copyrightyear{2026}
\acmYear{2026}
\setcopyright{cc}
\setcctype{by}
\acmConference[CHI '26]{Workshop paper submitted to 'Bidirectional Human-AI Alignment Workshop' at the 2026 CHI Conference on Human Factors in Computing Systems}{April 13--17, 2026}{Barcelona, Spain}
\acmBooktitle{Bidirectional Human-AI Alignment Workshop of the 2026 CHI Conference on Human Factors in Computing Systems (CHI '26), April 13--17, 2026, Barcelona, Spain}
\acmDOI{}
\acmISBN{}

%% file: ExSitu/paper-info.tex
\title{Design Generative AI for Practitioners: \\Exploring Interaction Approaches Aligned with Creative Practice}

\author{Xiaohan Peng}
\orcid{0000-0002-5202-419X}
\affiliation{%
  \institution{LISN\\ Université Paris-Saclay, CNRS, Inria
  }
  \city{Orsay}
  \country{France}
}
\email{xiaohan.peng@inria.fr}

\author{Wendy E. Mackay}

\orcid{0000-0001-8261-2382}
\affiliation{%
  \institution{LISN\\ Université Paris-Saclay, CNRS, Inria
  }
  \city{Orsay}
  \country{France}
    }
\email{wendy.mackay@inria.fr}
\author{Janin Koch}

\orcid{0000-0001-9207-9550}
\affiliation{%
  \institution{UMR 9189 CRIStAL\\ Univ. Lille, Inria, CNRS, Centrale Lille
  }
  \city{Lille}
  \country{France}
  }
  \email{janin.koch@inria.fr}

\input{sections/0_Abstract}

\ccsdesc[500]{Human-centered computing~Interactive systems and tools}
\ccsdesc[300]{Applied computing~Arts and humanities}

\keywords{Creativity Support Tools, Human-AI Interaction, Design Practice, Generative AI}

%% file: sections/0_Abstract.tex
\begin{abstract}
Design is a non-linear, reflective process in which practitioners engage with visual, semantic, and other expressive materials to explore, iterate, and refine ideas. 
As Generative AI (GenAI) becomes integrated into professional design practice, traditional interaction approaches focusing on prompts or whole-image manipulation can misalign AI output with designers’ intent, forcing visual thinkers into verbal reasoning or post-hoc adjustments. 
We present three interaction approaches from \textit{DesignPrompt}, \textit{FusAIn}, and \textit{DesignTrace} that distribute control across intent, input, and process, enabling designers to guide AI alignment at different stages of interaction. 
We further argue that alignment is a dynamic negotiation, with AI adopting proactive or reactive roles according to designers’ instrumental and inspirational needs and the creative stage.
\end{abstract}

%% file: sections/1_Introduction.tex
\section{Introduction}
Design is a ``reflective conversation with materials'' \cite{Schon2017:reflectivepractitioner}, where practitioners engage with visual, semantic, and other kinds of expressive mediums to guide their process. 
In professional practice, creation is rarely a linear path from concept to execution; rather, it is an entangled process of exploration, reflection, and refinement \cite{koch_semanticcollage_2020}. 
Designers build upon their work by reusing materials, maintaining precise control over details, and documenting the evolution of their ideas through sketches and annotations \cite{InieDalsgaard2020:manageidea, SCHON1992:Kindsofseeing}.

However, as Generative AI (GenAI) becomes more prevalent in design practice, its interaction paradigms often fail to adequately align designers’ expression and intent with the system’s responses.
The primarily conversation-driven interaction paradigms force visually oriented designers to shift from visual thinking to formulating verbal instructions \cite{koch_semanticcollage_2020, Park2024:WeAreVisualThinkers}.
Furthermore, there remains a conceptual gap between design vocabulary and GenAI terminology.
For instance, ``variation'' is an intentional iteration used to explore a design space for a designer \cite{logan1993:creativity}; while a variation is often merely a result of different random noise in a diffusion process for a GenAI system. 
Designers often have to settle on an initially plausible but less-than-ideal ``variation'', then make manual adjustments~\cite{Kocielnik2019:ImperfectAI}.

Existing visual interaction approaches primarily act on whole-image input, with optional masks to specify local detail control.
However, this interaction paradigm presupposes the existence of visual material that is already relatively complete. 
It forces designers to ``fix'' existing material rather than ``craft'' from scratch, shifting the focus from creative expression to post-hoc adjustment \cite{Vimpari2023:Adapt-or-Die}.
Moreover, design knowledge encompasses tacit abstractions, hierarchies, and practical constraints~\cite{son2024:demystifying} that are difficult to capture by executing text-to-image or image-to-image pipelines. 
The core issue is not the modalities GenAI affords nor how designers communicate, but how interaction is structured to elicit and capture design expressions, preserving meaningful design intentions, processes, and outcomes.

While some systems offer control through parametric components, these often demand technical expertise and lack intuitive correspondence to creative decisions \cite{Dang2022:GANSlider}. 
More advanced interfaces such as ComfyUI \footnote{https://www.comfy.org/} and Stable Diffusion WebUI \footnote{https://github.com/AUTOMATIC1111/stable-diffusion-webui} provide greater flexibility, but at the cost of further technical complexity. 
Design iteration is a purposeful, non-linear, and contextual process that cannot be adequately represented by one-dimensional parameters.
This highlights the need for interaction approaches that enable designers to guide AI alignment at their preferred level of abstraction without abandoning their workflows.

In this position paper, we present three different interaction approaches with GenAI applied to professional design practice showing the interaction diversity and design space that could be explored. 
We then reflect on the role of AI in different stages of human-AI collaborative design process.

%% file: sections/2_InteractionApproaches.tex
\section{Interaction Approaches for Creative Practice}
We argue that alignment between designers and GenAI is not a static goal achieved through optimizing any single point of the interaction --- whether input, output, or generation alone --- but emerges dynamically across the full arc of creative practice.
Current interaction approaches typically situate alignment at a single point, most often the prompt.
Yet professional design practice distributes intent, action, and reflection across time, space, and representation.
We developed three interaction paradigms that shift the locus of structure across the interaction: into the multi-modal decomposition of intent (\textit{DesignPrompt}), into the direct instrumental manipulation of visual prompts (\textit{FusAIn}), and into the spatial-temporal record of the design process itself (\textit{DesignTrace}). 
Together, they offer designers multiple points of leverage for steering GenAI's generative capabilities toward meaningful creative outcomes.

\subsection{Intent Alignment: Configurable Multimodal Scaffolds}
In recent HCI research, structuring and scaffolding GenAI input around design-relevant dimensions have emerged as an interaction approach for better expressing designer intent. 
Such input approach decompose the input into meaningful units such as design attributes, visual references, or semantic dimensions that designers can configure and recombine~\cite{DesignWeaver_Tao_25, Lu2025:Misty, CreativeConnect_Choi_24, ProductMeta_Zhou_25}.

\begin{figure}[h!]
    \centering
    \includegraphics[width=\linewidth]{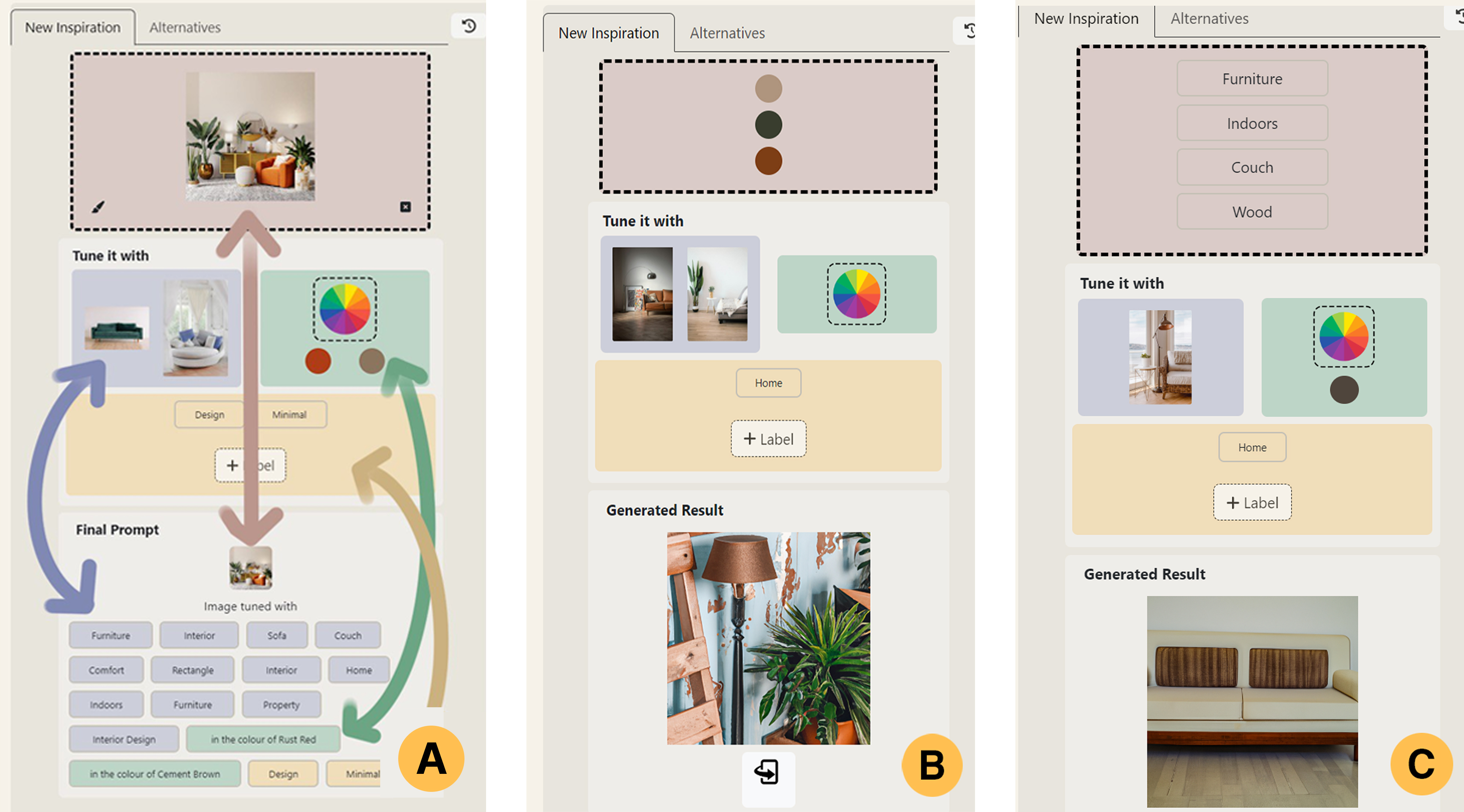}
    \caption{DesignPrompt supports intent decomposition through multiple input modalities: (A) image inpainting, (B) color, and (C) semantic tags. Multimodal inputs are mapped to editable text prompts, enabling interactive prompt editing and reordering to refine textual input to GenAI.}
    \label{fig:DesignPrompt}
    \Description{}
\end{figure}

One such example is DesignPrompt~\cite{peng2024:designprompt}, which lets designers decompose their intent into discrete and manipulable modalities: images, colors, and semantic tags (Fig. ~\ref{fig:DesignPrompt}). 
These composable elements form a multi-modal scaffold that makes the designer's visual intents explicit and individually adjustable. 
The system provides a transparent view of both the multi-modal input and the resulting translated prompt, enabling designers to refine their input before sending input to GenAI.
By separating intent into interactive components, designers can hold certain attributes such as a color palette constant, while exploring variation along others, enabling more precise alignment between what they envision and what the system generates.

\subsection{Input Alignment: Direct Manipulable Visual Prompts}
Alignment often breaks down when designers cannot intervene directly in the act of creation. 
Recent work has explored GenAI systems enabling practitioners to interact more actively and directly, moving beyond single-shot generation to interfaces where designers can iteratively shape visual outputs. 
Some systems address this by letting designers input at different levels of granularity~\cite{Sarukkai2024:BlockandDetail} and sketch continuously alongside high-fidelity generation~\cite{Inkspire_Lin_25, Liu2025:MagicQuill}.

\begin{figure}[h!]
    \centering
    \includegraphics[width=\linewidth]{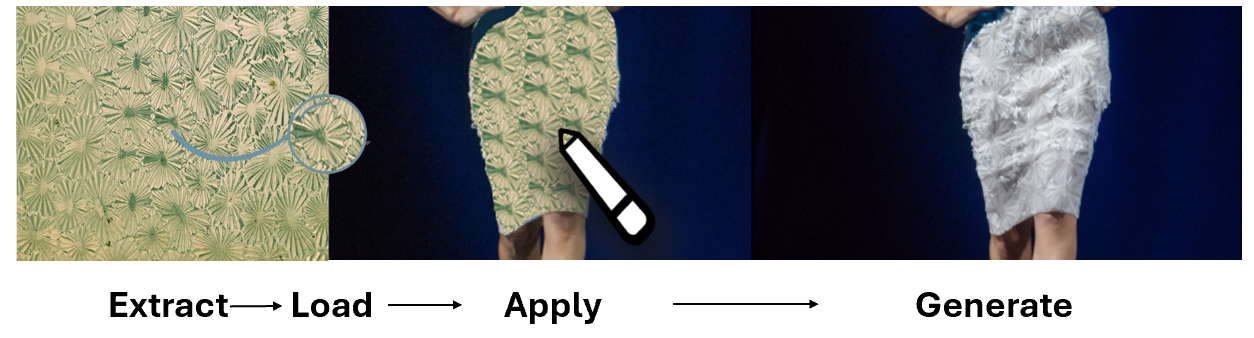}
    \caption{FusAIn reifies image-based visual prompts as personalized “pens” that encode specific visual properties. With “texture pens” users extract a texture from a source image, load it into a pen, and draw to apply that texture onto target visuals. The system then generates an image that preserves other properties while consistently applying the texture. 
    }
    \label{fig:FusAIn}
    \Description{}
\end{figure}

Our work on FusAIn~\cite{peng2025:FusAIn} extends this space by reifying text- and image-based prompts into instrumental ``smart'' pens that can be loaded with specific visual properties such as texture, color, or objects, then applied through direct manipulation on a canvas (Fig. ~\ref{fig:FusAIn}). 
This transforms the input from a static description into an expressive, spatial act: designers align the generation not by describing a desired result, but by physically constructing it through strokes and placement. 
The approach shifts interaction from a generate-and-verify loop to one of continuous, fine-grained visual guidance over details that are difficult to articulate through text or image alone.
\subsection{Process Alignment: Iterative and Trackable Generative Workflows}
Creative alignment between practitioners and GenAI is rarely achieved in a single generation; it emerges through cycles of non-linear exploration, comparison, and selective refinement that unfold over time. 
Yet such alignment is prone to cumulative generation drift, where repeated iterations with GenAI frequently diverge from the designer's original intent \cite{peng2026:designtrace}. 
Since each generation may involve ``destructive'' edits to previous outputs~\cite{Cao2025:Compositional}, failing to capture transitions between them makes it impossible to trace how creative decisions evolve, how alternatives relate, or why a particular direction was chosen.

\begin{figure}[h!]
    \centering
    \includegraphics[width=\linewidth]{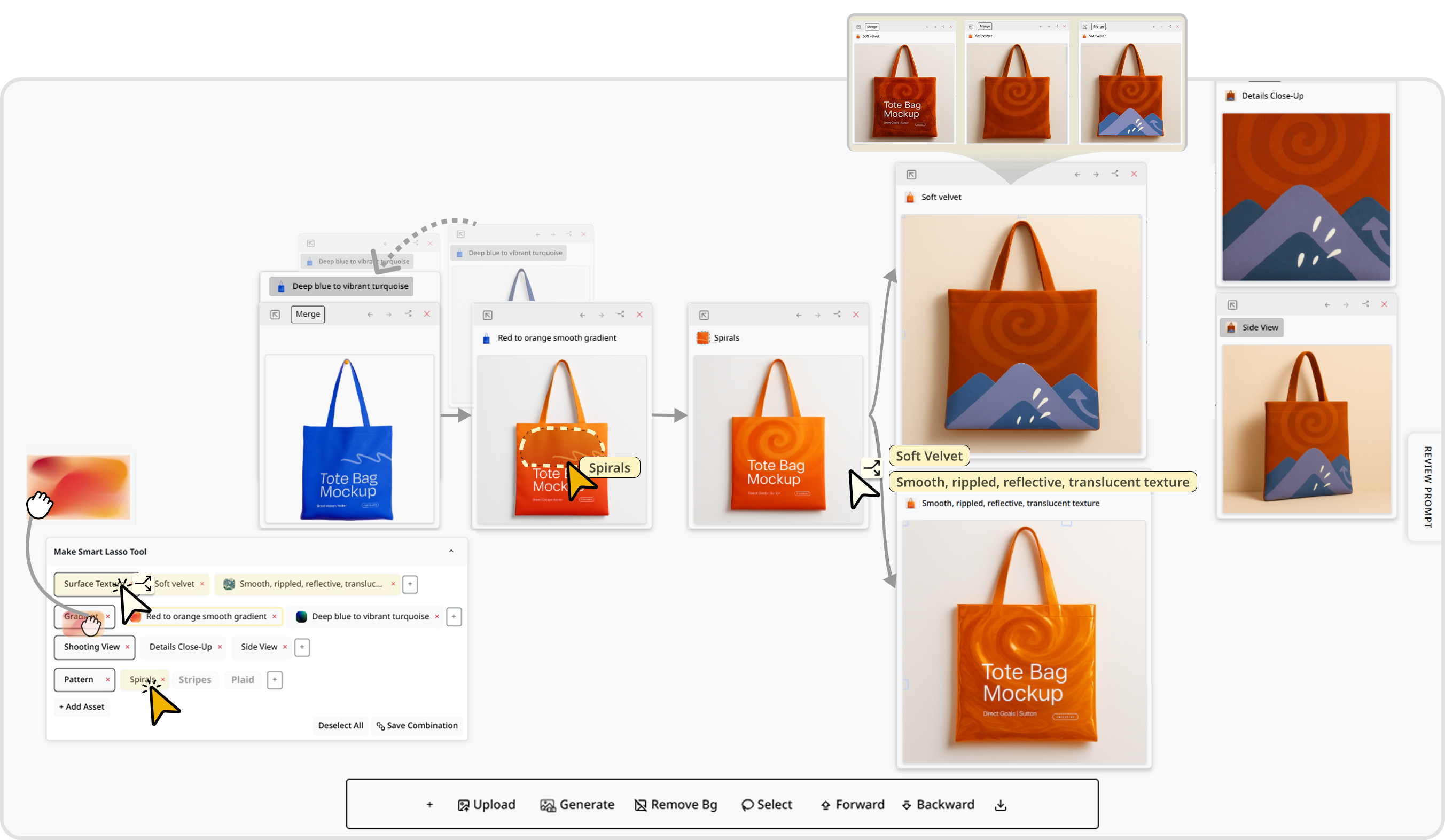}
    \caption{DesignTrace: a GenAI design tool for exploring alternatives and tracking design progress. Designers can extract semantic attributes from inspirational images and system suggestions, apply them to selected regions or entire images, branch multiple alternatives under shared semantic categories, customize prompts for image generation, make localized edits while navigating editing history, and collapse unused versions. 
    }
    \label{fig:DesignTrace}
    \Description{}
\end{figure}

DesignTrace \cite{peng2026:designtrace} introduces an iterative and trackable node-based interaction approach that captures and structures the traces of the non-linear design process with GenAI (Fig. ~\ref{fig:DesignTrace}). 
By reifying design rationale into visual and semantic artifacts and making these traces temporally and spatially trackable, the system helps designers maintain visual consistency across alternative explorations and iterations. 
Alignment is achieved through structural persistence: designers can revisit earlier states, branch new ideas without losing context, and observe the connections between iterations, ensuring the AI’s trajectory remains aligned with their long-term goals.

%% file: sections/3_RolesOfAI.tex
\section{Rethinking Roles of AI: Alignment as a Shifting Partnership}
The rapid integration of Generative AI (GenAI) into design practice calls for a shift from viewing AI merely as a passive tool to seeing it as a partner capable of shared agency. 
As ImageSense~\cite{koch_imagesense_2020} demonstrates, a collaborative human-AI ideation environment goes beyond passive search—it thrives on partnership, with shared agency and fluid roles. 
We therefore view ``alignment'' not as a fixed state of adherence to human instructions, but as a dynamic negotiation of initiative, shifting between leading and following depending on the designer's current cognitive state and the stage of the creative process~\cite{Lobo2024:When}. 

In human-human collaboration, leadership is rarely static; it evolves as partners construct a shared understanding of the problem. 
Similarly, AI agents must support different levels of initiative to act as effective partners. 

\subsection{Proactive AI: Designing with Productive Friction}
Alignment may benefit from ``productive friction''.
Drawing on the ``Shifting Initiative'' model \cite{Lobo2024:When}, the AI can take a leadership role, offering ``extreme suggestions'' or acting as a source of inspiration can broaden the idea space~\cite{koch_imagesense_2020}. 
Although high-initiative agents may sometimes be perceived as less ``warm'', they are valued for their efficiency, determination, and ability to drive the process forward when the user lacks a clear path~\cite{Lobo2024:When}.
The alignment in proactive AI-human interaction is about the AI understanding the breadth of the design space and taking initiative to provoke divergence rather than the precision of the output.

\subsection{Reactive AI: Designing with Instrumental Intent}
When designers have a clear instrumental intent, the AI may shift to a assistive ``Follower'' role. 
When designers know what they are looking for, they prefer tools that allow them to retain control, such as semantic search or human collaboration, rather than proactive AI suggestions. 
``Follower'' agents aligned their actions with the human's plan were perceived as warmer, more competent, and fostered stronger social identification with the team. 
When AI attempts to take initiative during execution-heavy phases, it risks being perceived as ``stubborn'' or disrupting the designer’s sense of control. 
The alignment in reactive AI-human interaction therefore requires the system to identify when to return control to the designer to ensure trust and user satisfaction.

\subsection{Shifting Initiative in Design Practice}
Designers valued in our work the ability to alternate between serendipitous exploration and focused work with the AI system depending on their immediate needs~\cite{koch_imagesense_2020}. 
Similarly, an AI partner must be capable of diagnosing the interaction context: knowing when to offer ``productive friction'' through leadership and when to strictly follow the designer’s intent. 
This fluid shifting of initiative mirrors the natural dynamics of human-human collaboration and supports meaningful alignment throughout the creative process.

%% file: sections/4_WSproposal.tex
\section{Conclusion}
We argue that alignment between designers and Generative AI is not a static objective achieved through more controllable prompts and outputs, but a multi-stage, multi-mode coordination that unfolds dynamically across the creative process.
Instead of optimizing alignment at singular points of interaction --- primarily through better prompting mechanisms or more steerable outputs, creative practice demands heterogeneous alignment.
By presenting three interaction approaches, we show alternatives how alignment can be continuously constructed and reconstructed through different modes of negotiation: 
Configurable multimodal scaffolds support alignment by structuring designers’ inputs;
Directly manipulable visual prompts enable alignment through direct visual negotiation; and 
Iterative, trackable workflows facilitate alignment by reifying design rationale and supporting process reflection.

The proposed multi-stage, multi-mode interaction approaches embrace the situated, non-linear nature of creative work --- where requirements emerge through doing, where ideas are progressively refined through remixing, and where alignment itself becomes part of the creative dialogue.
This multi-stage perspective reframes AI as a collaborative partner whose initiative and responsiveness can be dynamically coordinated across different phases of creative inquiry. 
We call for future GenAI systems that enable fluid shifts between proactive and reactive roles --- sometimes provoking divergence to expand the design space, sometimes executing refined intent with precision, and sometimes receding to preserve designer autonomy. 
The goal is to distribute structure and coordination across the workflow to support the evolving, reciprocal negotiation inherent in creative practice.